\documentclass[default]{aastex701}

\definecolor{dgreen}{RGB}{0,108,53}

\usepackage{adjustbox} 
\usepackage{booktabs}
\usepackage{float}
\usepackage{amsmath}
\usepackage{placeins}
\usepackage{tabularx}
\usepackage{xcolor}
\usepackage{dsfont}

\shorttitle{HelioFill: Far-Side EUV Estimation}
\shortauthors{Ben Ameur et al.}
\begin{document}

\title{HelioFill: History-conditioned Diffusion Estimation of Solar Far-side EUV Emission}

\author[0000-0001-5097-2885]{Firas Ben Ameur}
\email[show]{firas.benameur@kaust.edu.sa}
\affiliation{Computer, Electrical and Mathematical Sciences and Engineering Division, King Abdullah University of Science and Technology (KAUST) \\
23955-6900 Thuwal, Saudi Arabia}

\author[0000-0003-3670-4678]{Rayan Dhib}
\email{rayan.dhib@kuleuven.be}
\affiliation{Centre for Mathematical Plasma-Astrophysics, Department of Mathematics, KU Leuven \\
Celestijnenlaan 200-B, 3001 Leuven, Belgium}

\author[0009-0007-1706-013X]{Yahia Battach}
\email{yahia.battach@kaust.edu.sa}
\affiliation{Computer, Electrical and Mathematical Sciences and Engineering Division, King Abdullah University of Science and Technology (KAUST) \\
23955-6900 Thuwal, Saudi Arabia}

\author[0000-0003-4017-215X]{Andrea Lani}
\email{andrea.lani@kuleuven.be}
\affiliation{Centre for Mathematical Plasma-Astrophysics, Department of Mathematics, KU Leuven \\
Celestijnenlaan 200-B, 3001 Leuven, Belgium}

\author[0000-0001-7300-1280]{Matteo Parsani}
\email{matteo.parsani@kaust.edu.sa}
\affiliation{Computer, Electrical and Mathematical Sciences and Engineering Division, King Abdullah University of Science and Technology (KAUST) \\
23955-6900 Thuwal, Saudi Arabia}
\affiliation{Physical Science and Engineering Division, King Abdullah University of Science and Technology (KAUST) \\
23955-6900 Thuwal, Saudi Arabia}

\author[0000-0002-0973-1112]{Omar Knio}
\email{omar.knio@kaust.edu.sa}
\affiliation{Computer, Electrical and Mathematical Sciences and Engineering Division, King Abdullah University of Science and Technology (KAUST) \\
23955-6900 Thuwal, Saudi Arabia}

\author[0000-0002-1743-0651]{Stefaan Poedts}
\email{stefaan.poedts@kuleuven.be}
\affiliation{Centre for Mathematical Plasma-Astrophysics, Department of Mathematics, KU Leuven \\
Celestijnenlaan 200-B, 3001 Leuven, Belgium}
\affiliation{Institute of Physics, University of Maria Curie-Sk{\l}odowska \\
ul.\ Radziszewskiego 10, PL-20-031 Lublin, Poland}

\begin{abstract}
The loss of STEREO-B in 2014 created a persistent blind spot in extreme ultraviolet (EUV) imaging of the solar far-side. We present HelioFill, which is, to the authors' knowledge, the first denoising-diffusion inpainting model that estimates the unobserved sector of full-Sun 304\,\AA\ maps from the current Earth-side (Solar Dynamics Observatory, or SDO) and STEREO-A views. Because these views cannot constrain structures fully enclosed in the unobserved sector, HelioFill's diffusion stage is also conditioned on the sector's own rotation history, supplied as causal persistence and information-age maps; a rectified-flow refinement then adds realistic small-scale texture. Test rotations are separated from the diffusion stage's training timestamps by full-rotation guard bands. Over the unobserved sector, HelioFill's estimate correlates with the three-view truth at $\rho=0.520$, above persistence ($0.416$) and the input-independent climatological mean ($0.392$) on 97.9\% and 96.8\% of the test maps, which conditioning on the current views alone does not achieve; in peak signal-to-noise ratio (PSNR) it exceeds them by $2.27$ and $0.66$\,dB and matches a direct observation of the sector made $\approx$29\,hr before or after it ($\approx$49\,hr in correlation). The refinement lowers learned perceptual image patch similarity (LPIPS) by $11\%$ at a cost of $0.50$\,dB and $0.055$ in correlation. HelioFill localizes active regions better than persistence (Dice score $0.449$ versus $0.384$) and detects fully enclosed ones, although persistence detects more individual regions and coronal holes remain poorly reproduced. Against near-simultaneous far-side Solar Orbiter/EUI observations from 2022 to 2025, HelioFill correlates with the truth at $\rho=0.375$, above climatology ($0.286$) and level with persistence ($0.365$), which it exceeds in PSNR and structural similarity index (SSIM); its diffusion stage alone reaches $0.460$, above both, and the refinement stage costs the difference. HelioFill thus provides observationally plausible global EUV context within quantified limits.
\end{abstract}


\section{Introduction}\label{sec:intro}

Magnetic fields on the solar surface evolve continuously, shaping the large-scale magnetic structure of the corona and heliosphere~\citep{sheeley22005,wiegelmann2021solar}. Magnetically complex active regions (ARs) store magnetic energy that powers flares and coronal mass ejections (CMEs) that can disturb near-Earth space and impact ground-based and space-borne systems and infrastructures~\citep{leka2003photospheric,pal2018dependence}. The rapid emergence, evolution, and dispersal of ARs complicate analysis and forecasting, while the formation and stability of preeruptive magnetic structures remain incompletely understood~\citep{jarolim2024sunerf,jeong2025prediction}. Multiple space missions image the solar atmosphere in selected wavelength bands; in particular, extreme ultraviolet (EUV) imaging provides high-cadence, high-resolution views of the low corona, supporting the tracking of AR evolution and the identification of coronal holes.

During the 2010--2014 SDO--STEREO multiview interval, coordinated observations from the Atmospheric Imaging Assembly (AIA) aboard the Solar Dynamics Observatory (SDO; \citealt{pesnell2012solar,lemen2012atmospheric}) and the Extreme Ultraviolet Imager (EUVI) of the SECCHI suite \citep{howard2008secchi} aboard the twin STEREO spacecraft (\citealt{wuelser2004euvi}) provided multiview EUV coverage from the Earth line and from Earth-leading and Earth-trailing heliocentric viewpoints near 1~au~\citep{kaiser2008stereo}. When STEREO-A and STEREO-B approached $\pm90^{\circ}$ from the Sun--Earth line, and hence nearly $180^{\circ}$ from each other, near-simultaneous full-Sun EUV coverage became possible, enabling direct monitoring of far-side active regions before they rotated onto the Earth-facing disk. This overlap also supported high-quality multi-instrument products:~\citet{caplan2016synchronic} constructed synchronic EUV and coronal-hole maps at 6\,hr cadence (AIA~$193$ \AA\ with EUVI~$195$ \AA) using preprocessing that included PSF deconvolution, limb-brightening correction, and interinstrument intensity normalization, thereby improving the consistency of coronal-hole detection across viewpoints. The loss of routine contact with STEREO-B on 2014 October 1 ended the routine SDO--STEREO era of near-global multiview EUV context, leaving a persistent unobserved sector in global coronal monitoring.

Operational and research products therefore attempt to infer the far-side magnetic context indirectly, using synoptic or synchronic magnetic maps, helioseismic far-side detections, surface-flux-transport models, and, more recently, EUV-to-magnetogram inference. Synoptic magnetic maps stack Earth-facing magnetogram strips over a Carrington rotation, whereas synchronic maps evolve recently observed flux with surface-flux-transport models to provide full-surface boundary conditions for coronal and solar-wind models~\citep{bertello2014uncertainties}. For the far-side, data-assimilation frameworks can insert helioseismic far-side active-region detections \citep{arge2013adapt,yang2024combined} or magnetic fields inferred from them, or active regions identified in STEREO 304\,\AA\ images \citep{upton2024aft}, into surface-flux-transport (SFT) models~\citep{devore1984concentration,schrijver2003photospheric}. Helioseismic imaging locates the largest and brightest far-side active regions reliably \citep{liewer2014testing,yang2023direct}. However, helioseismic far-side imaging remains limited in spatial detail and is less reliable for small, weak, or short-lived emergences. Standard SFT models cannot reliably introduce newly emerging far-side flux unless it is supplied by observations or proxy detections, a limitation that is particularly acute for regions emerging or evolving while outside the Earth-facing disk. Consequently, these methods improve global magnetic context but do not restore the fine-scale EUV morphology needed for continuous full-Sun coronal monitoring~\citep{yang2024combined}.

Data-driven models have begun to supplement synoptic and SFT pipelines by extracting magnetic-field information directly from EUV imagery. \citet{felipe2019improved} trained a convolutional network on far-side helioseismic holography phase-shift maps, calibrated against delayed Helioseismic and Magnetic Imager (HMI) magnetograms, to produce probability maps of far-side ARs presence. Their approach increases sensitivity relative to standard holography; however, the resulting maps remain coarser than direct EUV imagery and do not provide magnetic polarity information. \citet{jeong2020solar} trained a conditional model to translate three-channel EUV images into HMI-like magnetograms, showing strong agreement in global and regional flux metrics and establishing EUV$\rightarrow$magnetogram inference as a viable proxy pathway. Extending to the far-side, \citet{jeong2022improved} produced ``AISFM~3.0'' far-side magnetograms from STEREO/EUVI together with front-side reference data (2011--2021), reporting improved pixel-level correlations and structural similarity, as well as consistent polar-field trends for solar cycles 24 and 25. Related work has produced far-side magnetograms from STEREO/EUVI images with deep networks \citep{kim2019farside}, reconstructed far-side vector magnetograms from EUV data and an SFT field propagated from one rotation earlier \citep{jeong2025farside}, and converted helioseismic images into far-side flux maps \citep{chen2022farside}. More recently, \citet{jarolim2025deep} introduced a deep instrument-to-instrument (ITI) translation framework that homogenizes long multimission EUV archives (e.g., EIT/EUVI$\rightarrow$AIA) and can also approximate unsigned magnetic-field estimates from EUV. \citet{jarolim2024sunerf} adapt neural radiance fields to the optically thin corona, learning a 3D, time-aware EUV radiance field from AIA+EUVI that supports novel view rendering and height estimation of coronal structures. This capability is powerful, but computationally more expensive than cadence-driven operational mapping. EUV-driven forecasting has also matured; for example, deep models using AIA~193/211\AA\ outperform WSA--ENLIL baselines for near-Earth wind speed prediction in high-speed stream regimes, underscoring the operational value of EUV-based inference when appropriately trained and validated~\citep{son2023three}.

These developments motivate a fast image-domain estimator of the missing far-side view at operational cadence. Denoising diffusion probabilistic models have become a leading class of generative models for image synthesis~\citep{ho2020denoising} and have been applied to SDO/AIA imagery \citep{ramunno2024solar}, and related diffusion-based inpainting methods have shown strong performance in reconstructing missing or masked image regions~\citep{rombach2022high,lugmayr2022repaint}. Their iterative refinement process provides a natural mechanism for combining local texture recovery with global structural consistency, two requirements that are particularly important when filling large gaps in scientific images. PixelHacker~\citep{xu2025pixelhacker}, a recent diffusion-based inpainting framework, further highlights the importance of preserving structural and semantic consistency under challenging masks. In parallel, foundation models pretrained on multichannel SDO observations provide transferable solar representations across tasks such as active-region segmentation, flare forecasting, and solar-wind prediction, suggesting a promising path toward probabilistic, physics-aware generative models for solar image reconstruction~\citep{roy2025surya}.

In this Letter, we bring recent advances in diffusion-based inpainting to solar physics. Specifically, we adapt a modern inpainting framework, PixelHacker, to the EUV setting and introduce HelioFill, a mask-conditioned model for full-Sun Carrington cylindrical equal-area (CEA) maps. Trained on maps from 2011 to 2014, HelioFill learns to estimate the unobserved far-side region resulting from the loss of STEREO-B, using Earth-side and STEREO-A views as input.

Physically, the far side holds two kinds of structure that the available data constrain differently. We distinguish throughout between \emph{reconstruction} of boundary-supported structure, which the visible context constrains, and \emph{generation} of a plausible realization of fully enclosed structure, which it cannot determine uniquely. The best predictor of an enclosed structure is its appearance at its last observation, which no current image contains, so HelioFill also conditions on the sector's rotation history, supplied as a persistence map and a per-pixel information-age map. This input is related to the SFT prior of \citet{jeong2025farside}, a modeled field propagated from magnetograms one rotation earlier, but is an observed image, typically about a day old on the test rotations, and censored against the held-out rotations during training. The two inputs are complementary: the current views constrain boundary-supported structure, whereas the rotation history provides the information on fully enclosed structure that no current image contains. A latent-diffusion model produces the estimate, and a rectified-flow refinement adds realistic small-scale texture. We evaluate HelioFill under a temporally blocked protocol against nonlearning baselines, measure the spatial scales at which it carries positional information, and validate it against near-simultaneous far-side observations by Solar Orbiter.

Filling the far-side provides plausible active-region context and limb continuity, complements helioseismic/SFT far-side estimates, and could provide cleaner inner-boundary conditions for coronal and heliospheric models, whose solutions are sensitive to far-side structure \citep{perri2024impact}, when assets are lost or intermittent.

\section{Data and Evaluation Protocol}\label{sec:methods}

\subsection{Dataset Description}\label{sec:data}

We construct a training dataset from the 2011--2014 interval when SDO/AIA and both STEREO/EUVI spacecraft provided nearly continuous 360$^{\circ}$ coverage. AIA observations were processed to Level~1.5 using the \texttt{aiapy} calibration tools~\citep{Barnes2020}, which update pointing information and register the images to a common plate scale. EUVI observations from both STEREO spacecraft were calibrated using the \texttt{secchi\_prep} routine in SolarSoft~\citep{SOLARSOFT}. HelioFill uses the 304~\AA\ band, which highlights filaments, active regions, and coronal holes.

For each valid time with simultaneous AIA, EUVI-A, and EUVI-B observations, we generated Carrington cylindrical equal-area (CEA) longitude--latitude mosaics using SunPy's reprojection framework~\citep{sunpy_community2020}. Prior to reprojection, each image was clipped to the visible solar disk, scaled by the median on-disk intensity to reduce exposure- and instrument-dependent variations, and transformed with a logarithmic stretch. A cosine-based weighting was then applied so that disk-center pixels contribute more strongly than near-limb pixels. The weighted images were reprojected and coadded to produce $1024 \times 2048$ synchronic maps.

\begin{figure}[!t]
\centering
\includegraphics[width=\textwidth]{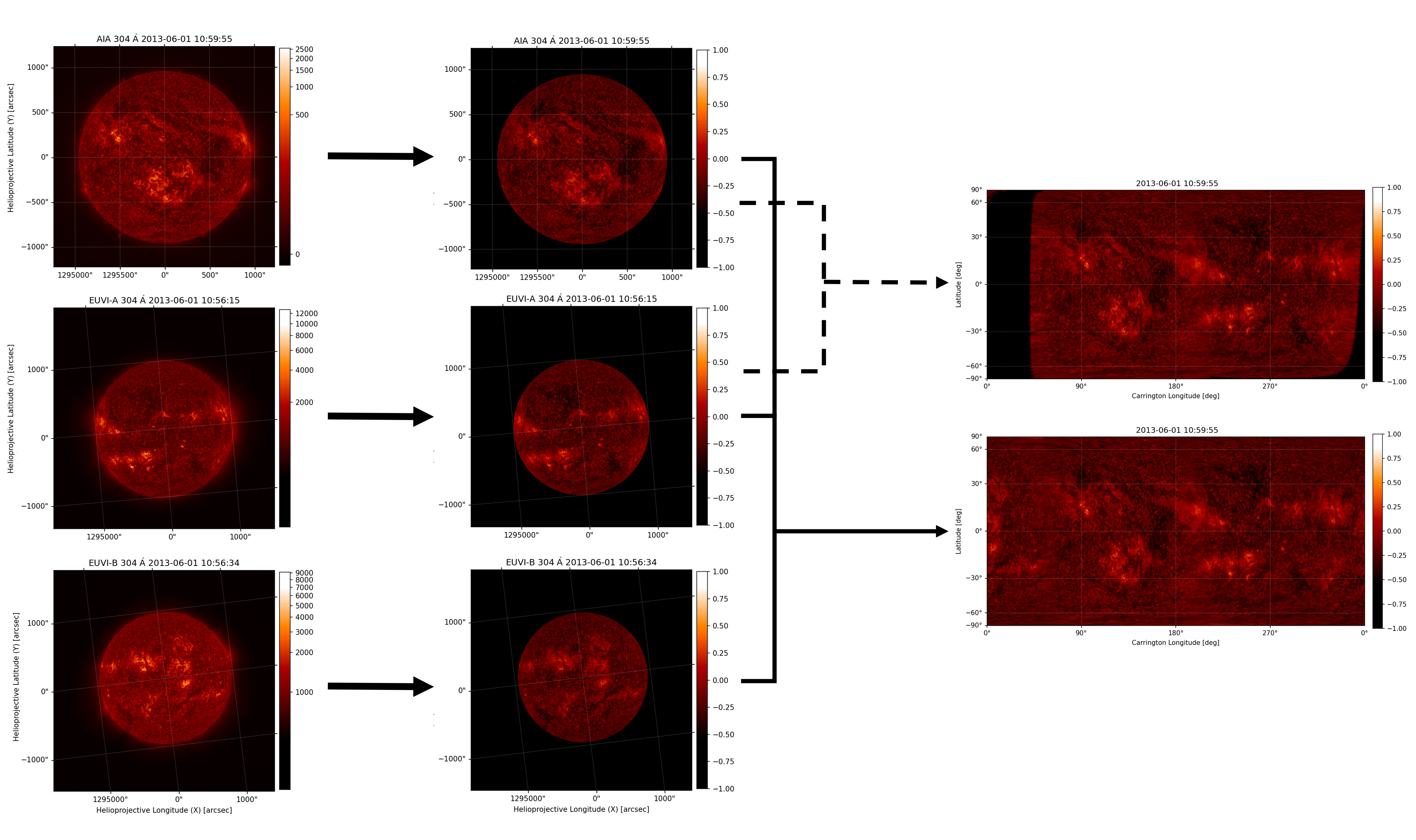}
\caption{Illustration of the dataset construction workflow. \textbf{Left column:} raw 304~\AA\ observations from SDO/AIA and STEREO/EUVI-A/B. \textbf{Middle column:} preprocessed images after calibration, disk clipping, intensity rescaling, and normalization. \textbf{Right column:} CEA synchronic maps. The upper map shows the input configuration with the STEREO-B coverage gap masked (used as model input), while the lower map shows the corresponding full three-view mosaic serving as the target ground truth.}
\label{fig:dataset}
\end{figure}

To emulate post-2014 observing conditions, when STEREO-B data were unavailable, we constructed input-target pairs by masking the far-side sector that would have been observed only by STEREO-B. The Earth-side and EUVI-A views were retained as model input, while the corresponding three-view mosaic was used as the ground-truth target. Pixels in the masked region were replaced by a constant fill value of $-100$, and binary masks were generated to explicitly identify the missing areas. All mosaics were downsampled to $512 \times 1024$ pixels for computational efficiency while preserving the global solar morphology. Approximately $10{,}000$ paired examples were produced for model training and evaluation; Figure~\ref{fig:dataset} illustrates the processing pipeline.

To augment the training set beyond its 4898 real maps (Section~\ref{sec:train}), we generate $N=10$ masked variants for each real mosaic, always using the unmasked three-view mosaic as the target. Masks are applied in CEA coordinates and include two classes: (i) \emph{geometry-aware} occlusions mimicking observational constraints (longitudinal wedges of 70--140$^{\circ}$ that spare the polar caps, bands at the longitude edges of the map, and longitudinal telemetry gaps); and (ii) \emph{random} rectangles, stripes, and circular cutouts. Each variant inherits its parent's data subset and history input.

We denote by $S$ the \emph{unobserved sector}, the pixels observed by neither SDO nor STEREO-A, and by $M\supset S$ the \emph{masked region} that HelioFill treats as missing: $S$ widened by 20\,px ($7^{\circ}$) in longitude on each side (not across the $0^{\circ}/360^{\circ}$ edge of the map), a margin in which the two-view composite and the three-view reference already disagree. On average 28\% of $M$ lies outside $S$.
\subsection{Temporally Blocked Splits}\label{sec:train}

Consecutive maps are nearly identical and long-lived structures persist for days, so a random split by timestamp leaks near-duplicate solar states into evaluation; Appendix~\ref{app:leakage} quantifies this leakage for such a split of our archive. We therefore hold out complete Carrington rotations: CR~2115, 2128, 2141, and 2151, one per year, form the test set, CR~2122 and 2138 the validation set, and the rotation on each side of every held-out rotation is a guard band whose maps are not diffusion training samples. The minimum separation between training and test times is 27.3\,days. The split leaves 4898 training, 645 validation, and 1343 test timestamps; the 3222 guard timestamps serve as training samples only for the refinement network, which excludes those within 7\,days of any validation or test timestamp (Section~\ref{sec:pmrf}), and for the VAE below. Thirteen test maps have no usable truth over $S$ and are excluded from all test statistics: four whose three-view target is exactly zero over $S$, and nine taken when an input view is missing or incomplete, whose target is zero over 33\%--81\% of $S$. Checkpoints and inference settings are selected on the validation rotations, and for each diffusion model the validation-selected checkpoint also has the highest test peak signal-to-noise ratio (PSNR) over $M$, the selection criterion, among its checkpoints scored on the test set; Appendix~\ref{app:pmrf} lists further test-set evaluations. The frozen variational autoencoder (VAE) of Section~\ref{sec:diffusion} predates the split: it was trained on a seeded random 90\% of the 10{,}108 three-view targets, including most of the test-rotation targets; it compresses each map independently, and Appendix~\ref{app:bands} bounds what its latent representation carries.

\subsection{Baselines, Control, and Metrics}\label{sec:exp-design}

A generative estimator can score well by exploiting temporal persistence, memorized configurations, or its learned image prior rather than the observations. We compare HelioFill with baselines and a control that isolate these mechanisms:
\begin{enumerate}
    \item \textit{Persistence.} For each pixel of $M$, the most recent direct two-view observation strictly before the sample time (per-map mean information age over $M$: median 1.2\,days, mean 2.2\,days), the same history that HelioFill receives.
    \item \textit{Nearest-neighbor retrieval.} The target of the training sample whose observed region best matches, pasted into $M$; it represents reproduction of a single memorized configuration.
    \item \textit{Climatology.} The pixel-wise mean of the training targets, the input-independent predictor with the lowest expected squared error on the training distribution.
    \item \textit{Prior-only control.} HelioFill's diffusion stage trained without the history input (the \emph{Without history} ablation of Table~\ref{tab:main}), sampled with its two-view input and mask blanked, so that the fill comes from the learned prior alone.
\end{enumerate}
All diffusion models use the same training loop, optimizer, and diffusion schedule, and all experiments the same evaluation protocol. We first retrain PixelHacker~\citep{xu2025pixelhacker} on this dataset to establish a baseline that reflects the original architecture: a conditional UNet with Gated Linear Attention (GLA) trained with uniform timestep weights, no confidence weighting, and $\beta=1$ in Equation~(\ref{eq:wspatial}). HelioFill uses the same VAE front end and data preprocessing, and we progressively enable its additional components to isolate their individual impact. Ablations disable the components of HelioFill cumulatively, from the full model down to the retrained PixelHacker. Table~\ref{tab:main} names each rung by what it keeps: \emph{Diffusion stage, harmonized sampling} is HelioFill without its refinement stage; \emph{diffusion stage, standard sampling} further gives up harmonized sampling (Section~\ref{sec:harmonized}); \emph{without history} further removes the history channels; \emph{retrained PixelHacker} further removes the solar-specific modifications, namely loss weighting, training on $M$ rather than on $S$, and training-time spectral gating, and is the retrained PixelHacker described above. Each variant is trained from scratch under identical data, budget, and hardware, with validation-based checkpoint selection, ensuring that measured differences stem from the activated components; each variant is a single training run.

Over $S$ we report the Pearson correlation $\rho$ between estimate and target, which measures whether structure is in the right place irrespective of its amplitude, and PSNR \citep{jarolim2025deep} over $S$ and over $M$. Because squared error rewards an estimate whose contrast, the ratio of its standard deviation to the target's over $S$, is reduced in proportion to its correlation with the target, PSNR favors estimates of reduced contrast, and a smooth estimate can outscore a better-placed one; $\rho$ is unaffected by contrast.

We also assess fidelity with the structural similarity index (SSIM; \citealt{wang2004image}) over $S$, a seam statistic (the absolute difference between the squared predicted and true jumps in mean intensity across the boundary of $M$, each jump measured between 4\,px rings on either side), and learned perceptual image patch similarity (LPIPS; \citealt{zhang2018unreasonable}), computed on the bounding box of $M$ padded by 8\,px. For structure, active regions (ARs) and coronal holes (CHs) are the pixels brighter than the 95th, or darker than the 10th, percentile of the pixels outside $M$ of the map being segmented (target and estimate each on its own), compared inside $M$ by Dice score, false-alarm rate (the fraction of predicted components without a true counterpart), and detection rate (the fraction of true components with at least 30\% of their area predicted), with components smaller than 64\,px ignored and true components split into those that touch the edge of $M$ or the map border and those that do not. Paired differences are quoted as mean $\pm$ 1.96 standard errors over the test maps, treating the maps as independent.

\section{Model}\label{sec:model}

\subsection{Overview}

\begin{figure}[!t]
\centering
\includegraphics[width=\textwidth]{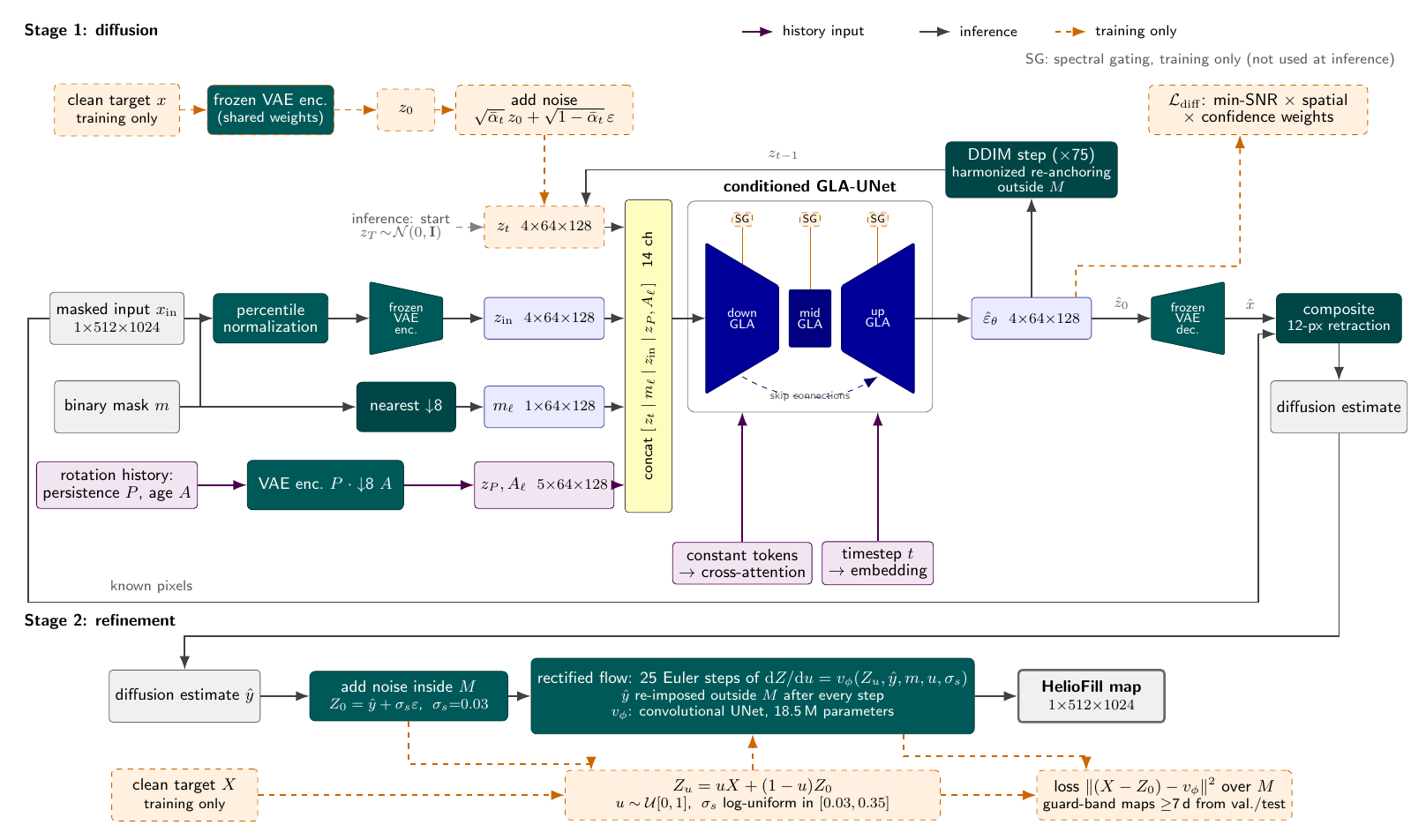}
\caption{HelioFill. Top: diffusion stage. The masked two-view input, the persistence map $P$, and the information-age map $A$ are brought to the latent grid and concatenated with the noised latent $z_t$ and the mask (14 channels); a GLA-UNet with constant learned context tokens predicts the noise, and 75 DDIM steps with harmonized reanchoring yield the diffusion estimate. Bottom: refinement stage, a rectified flow that adds texture inside the masked region. Black: inference path; dashed orange: training only; violet: history input.}
\label{fig:arch}
\end{figure}

HelioFill estimates $M$ in two stages (Figure~\ref{fig:arch}). A latent-diffusion model produces an estimate $\hat{y}$ conditioned on the observed pixels and on the sector's rotation history, and a refinement network then adds realistic small-scale texture to $\hat{y}$. Observed pixels farther than 32\,px ($11^{\circ}$) in longitude from $S$ are copied from the input. The diffusion stage is a latent-diffusion inpainting model tailored to single-channel solar EUV CEA maps at a resolution of $H \times W = 512 \times 1024$. First, a custom VAE with $\times 8$ downsampling, four latent channels, and latent scaling maps each input image to a compact latent representation $\Omega_\ell \in \mathbb{R}^{4 \times 64 \times 128}$, where diffusion can be performed efficiently~\citep{rombach2022high}. Second, a conditioned UNet denoiser, augmented with GLA~\citep{yang2023gated,xu2025pixelhacker} and training-time spectral-gating blocks~\citep{roy2025surya}, performs mask-aware denoising directly in this latent space. Training minimizes a min-SNR-weighted $\epsilon$-prediction loss~\citep{ho2020denoising,hang2023efficient}, with additional spatial and confidence weights whose product concentrates the learning signal on the masked region $M$.

\subsection{Mask-aware Standardization and VAE Interface}\label{sec:diffusion}
Let $x\in\mathbb{R}^{1\times H\times W}$ denote the input CEA image and $m\in\{0,1\}^{1\times H\times W}$ a binary mask with $m{=}0$ on the masked region $M$ (Section~\ref{sec:data}) and $1$ elsewhere. 
We compute per-sample percentiles $(p_{\mathrm{low}},p_{\mathrm{high}})$ as the 1st and 99th percentiles of the three-view target, which therefore depend on the unobserved sector, and normalize the image $x$ using Equation~\eqref{eq:norm}, 
\begin{equation}
\label{eq:norm}
\hat x=2\cdot\min\left(\max(\frac{x-p_{\mathrm{low}}}{p_{\mathrm{high}}-p_{\mathrm{low}}},0),1\right)-1.
\end{equation}
Unobserved pixels in $x$ (the $-100$ fill) are filled with $p_\mathrm{low}$ to avoid introducing a scale bias; observed pixels of $M$ keep their values. Appendix~\ref{app:leakage} shows that computing the bounds from observed pixels alone changes HelioFill's PSNR over the unobserved sector by only $-0.03\pm0.01$\,dB, and the Solar Orbiter evaluation of Section~\ref{sec:so} uses observed pixels throughout. The normalized image $\hat{x}$ is then encoded through the frozen VAE using the posterior mean, yielding the latent representation
\begin{equation*}
    z_{\mathrm{in}} = s\,\mathrm{VAE}_{\mathrm{enc}}(\hat{x}),
\end{equation*}
where $s=0.13025$ is the latent scaling factor; the three-view target, normalized with the same bounds, is encoded in the same way to the target latent $z_0$. The latent mask $m_\ell$ is obtained by nearest-neighbor downsampling of $m$, preserving its binary support. The VAE remains frozen during diffusion training.

\subsection{Diffusion Forward Process and Denoiser Input}
Given a Denoising Diffusion Probabilistic Model (DDPM) noise schedule $\{\alpha_t\}_{t=1}^T$ (Appendix~\ref{app:model}) and its cumulative product $\bar{\alpha}_t=\prod_{s=1}^{t}\alpha_s$, we draw $\varepsilon\sim\mathcal{N}(0,\mathbf{I})$ and construct the noised latent
\begin{equation}
\label{eq:forward}
z_t = \sqrt{\bar{\alpha}_t}\, z_0 + \sqrt{1-\bar{\alpha}_t}\,\varepsilon .
\end{equation}
The UNet receives the latent inpainting input $[z_t, m_\ell, z_{\mathrm{in}}, z_P, A_\ell]$, which contains 14 channels: four from the noised latent $z_t$, one from the latent-resolution mask $m_\ell$, and four from the encoded masked observation $z_{\mathrm{in}}$, plus the five history channels $z_P$ and $A_\ell$ of Section~\ref{sec:history}. Here, $z_t$ is the noised version of the target latent, $m_\ell$ identifies the missing region, and $z_{\mathrm{in}}$ provides contextual information from the masked observation.

\subsection{Rotation-history Conditioning}\label{sec:history}

The current views, those taken at the sample time, cannot constrain structures fully enclosed in the unobserved sector, but the sector's recent history can. HelioFill therefore receives a persistence map $P$, holding for every pixel the most recent directly observed two-view value strictly before the sample time, built by one chronological pass over the observation archive without reading any target mosaic, and an information-age map $A$, the time elapsed since that observation (per-map mean over $M$: median 1.2\,days on the test rotations). $P$ is normalized like the input and encoded by the VAE (four channels); $A$ is scaled as $\min(\mathrm{age}/10\,\mathrm{days},\,1)$ and downsampled bilinearly (one channel). Because persistence approximates the held-out truth, this input would leak evaluation data unless censored. During training, the buffer that produces $P$ never absorbs observations made within a held-out rotation (the validation buffer likewise skips the test rotations), so censored pixels keep an older value and a correspondingly larger age, as after an operational data gap; a check of every training snapshot finds no violation. Observations made during the guard rotations are not censored and can therefore enter the history of a training map. At test time the buffer is uncensored, as for the persistence baseline.

\subsection{Conditioned UNet with Spectral Gating}
The conditioned UNet with GLA blocks predicts the noise estimate
\(\hat\varepsilon_\theta=\mathrm{UNet}([z_t,m_\ell,z_{\mathrm{in}},z_P,A_\ell],t,\mathbf{C}), \quad \mathbf{C}\in\mathbb{R}^{B\times T\times D}\).
Here, $\mathbf{C}$ denotes a token-sequence context. For each batch of size $B$, we deterministically construct a per-sample sequence of length $T$$=10$ from the constant token indices $0,\dots,9$, identical for every sample at training and inference and carrying no metadata of any kind. These token IDs are mapped to $D$-dimensional vectors through a learned embedding table and injected into the UNet through cross-attention. The UNet has block widths 320, 640, 1280, and 1280, GLA replacing self-attention, and timestep embeddings added in the residual blocks. The widths follow the Stable Diffusion UNet layout inherited from PixelHacker: they double from the first to the third resolution level and are held at 1280 at the fourth, deepest level rather than doubling again; that level's down and up blocks are residual only, and the middle block between them carries GLA and cross-attention.

A clean latent estimate is recovered as
\begin{equation}
\label{eq:pred}
\hat z_0 \;=\; \frac{z_t \;-\; \sqrt{1-\bar\alpha_t}\,\hat\varepsilon_\theta}{\sqrt{\bar\alpha_t}}.
\end{equation}

To improve global structure and texture reconstruction, we insert spectral-gating modules during training in the second downsampling block, the middle block, and the second upsampling block. Each module applies a real 2D Fast Fourier Transform (FFT), derives channel-wise weights from the magnitude spectrum using a multilayer perceptron (MLP), modulates the complex-valued spectrum with these weights, and maps the result back to the spatial domain through an inverse FFT; the result, scaled by 0.1, is added to the block output, and the branch is dropped at inference. This yields a lightweight per-channel modulation mechanism~\citep{suvorov2022resolution,roy2025surya}.

\subsection{Losses}
The base prediction loss is defined as
\begin{equation}
\label{eq:lpred}
 \mathcal{L}_{\mathrm{pred}} = \|\hat{\varepsilon}_\theta-\varepsilon\|_2^2.
\end{equation}
We use min-SNR temporal weighting, $w_t$, with $\gamma=4.0$ to balance early versus late timesteps,
\begin{equation}
\label{eq:wt}
 w_t = \frac{\min(\gamma,\mathrm{SNR}_t)}{\mathrm{SNR}_t}, \quad \mathrm{SNR}_t = \frac{\bar{\alpha}_t}{1-\bar{\alpha}_t}.
\end{equation}
A spatial weighting, $w_{\mathrm{spatial}}$, emphasizes masked-region reconstruction and reads
\begin{equation}
\label{eq:wspatial}
 w_{\mathrm{spatial}} = 1 + \beta(1-m_\ell), \quad \beta=6.0.
\end{equation}
In addition, we use a confidence weighting, $w_{\mathrm{conf}}$: 
\begin{equation}
w_{\mathrm{conf}} = (1-\alpha) + \alpha \cdot \text{softmax}(c_{\text{base}}^{\eta}/\tau) \cdot N,
\label{eq:wconf}
\end{equation}
where $\eta=1.2$ controls emphasis on high-confidence regions, $\tau=1$ is the softmax temperature, $\alpha=0.5$ controls the blend with uniform weighting, and $N$ is the number of latent elements per sample, so that the temperature-scaled softmax (taken per sample over all latent elements) has unit mean. The confidence map, $c_{\text{base}}$, is the indicator of observed pixels outside $M$, downsampled to the latent grid,
\begin{equation}
c_{\text{base}} = \mathcal{D}_8\!\left(1-\mathds{1}_M\right),
\label{eq:cbase}
\end{equation}

where $\mathds{1}_M$ is the indicator of $M$ and $\mathcal{D}_8$ antialiased bilinear downsampling by a factor 8, so that $c_{\text{base}}$ is 0 on $M$, 1 on the observed latents, and intermediate only at the edge of $M$ (up to subcell differences where the edge of the missing region runs along latitude or meets the $0^{\circ}/360^{\circ}$ edge of the map). Because the softmax favors observed latents by up to a factor $e$, $w_{\mathrm{conf}}$ partly offsets $w_{\mathrm{spatial}}$.

The complete weighted diffusion loss combines spatial, temporal, and confidence weighting as follows:
\begin{equation}
\label{eq:ldiff}
 W = w_{\mathrm{spatial}} \cdot w_t \cdot w_{\mathrm{conf}}, \quad \mathcal{L}_{\mathrm{diff}} = \frac{\langle W, \mathcal{L}_{\mathrm{pred}}\rangle}{\|W\|_{1}}.
\end{equation}
$\mathcal{L}_{\mathrm{diff}}$ is the training objective; all parameter updates are computed from this term alone.

\subsection{Inference and Composition}
For inference, we use EMA-averaged (exponential moving average) weights and deterministic Denoising Diffusion Implicit Model (DDIM) sampling with 75 denoising steps \citep{song2021ddim} and a fixed seed. At each step, until the release of Section~\ref{sec:harmonized}, the latent outside $M$ is replaced by the encoded observation, a noise-free latent-space variant of the per-step known-region replacement of RePaint \citep{lugmayr2022repaint}; after the last step, the decoded estimate is composited with the observed pixels. Observed pixels are copied from the standardized input beyond 12\,px from $M$ (Section~\ref{sec:harmonized}), without feathering. Inference on a single NVIDIA A100 80\,GB GPU is reproducible under fixed settings and fast enough to support near-real-time operational pipelines.

\subsection{Harmonized Sampling and Boundary Extension}\label{sec:harmonized}

Two inference-time modifications further improve operational estimates. First, in the baseline inpainting sampler, the known-region latents are reclamped to the encoded input at every DDIM step. This replacement can force the denoiser to reconcile a clean known region with a still-noisy masked region. We therefore release this clamp during only the final 30 of the 75 DDIM steps, allowing the known and generated regions to codenoise near the end of sampling. Second, the two-view input differs from the three-view reference most strongly near the limb, where the missing third view would otherwise contribute. We therefore place the composite copy boundary 12 pixels in longitude inside the observed region and allow the harmonized reconstruction to fill the intervening band. The number of released steps was chosen on the validation rotations from $\{0,5,10,20,30,45,60\}$ by PSNR over $M$, which peaks at 30; release alone worsens the seam, and the retraction restores it. The combined setting was required, under a criterion fixed in advance, to improve PSNR over $M$, the seam statistic, and PSNR over the observed region together relative to full reanchoring, which it does on the validation rotations ($15.75\to15.94$\,dB, a fourfold lower mean seam, and $24.64\to25.27$\,dB). Retractions of 8, 12, and 16\,px gave identical validation PSNR over $M$ and mean seams within 0.2\%, and we adopted the intermediate value.

\subsection{Refinement Stage}\label{sec:pmrf}

Trained with a squared-error objective and sampled deterministically, the diffusion stage yields an estimate of the unobserved sector that is smoother than the observed maps. HelioFill therefore ends with a refinement stage that keeps the deterministic estimate and adds small-scale texture generatively, the predict-and-refine design of \citet{whang2022deblurring}. A rectified flow \citep{liu2023flow} transports the noised estimate $Z_0=\hat{y}+\sigma_s\varepsilon$, with $\varepsilon\sim\mathcal{N}(0,\mathbf{I})$ and $\sigma_s$ in the units of Equation~(\ref{eq:norm}), toward the target $X$ along straight paths $Z_u=uX+(1-u)Z_0$, $u\in[0,1]$, inside $M$, a transport that starts from an image rather than from noise, as in inversion by direct iteration \citep{delbracio2023inversion}. A velocity network $v_\phi(Z_u,\hat{y},m,u,\sigma_s)$, with $m$ the mask of $M$, is trained with the loss $\|(X-Z_0)-v_\phi\|_2^2$ over $M$, and inference integrates $\mathrm{d}Z/\mathrm{d}u=v_\phi$ from $Z_0$ in 25 Euler steps with $\sigma_s=0.03$, reimposing $\hat{y}$ outside $M$ after every step. The 12\,px band between $M$ and the copied observations therefore keeps the diffusion-stage estimate and is visible as a smoother strip 20--32\,px from the sector. The network, a convolutional UNet described in Appendix~\ref{app:pmrf}, is trained on diffusion-stage estimates and targets for the guard-band timestamps at least 7\,days from any validation or test timestamp, and its checkpoint and operating point are chosen on the validation rotations (Appendix~\ref{app:pmrf}).

\section{Results}\label{sec:results}

\subsection{Agreement Over the Unobserved Sector}\label{sec:fidelity}

\begin{figure}[!t]
\centering
\includegraphics[width=0.92\linewidth]{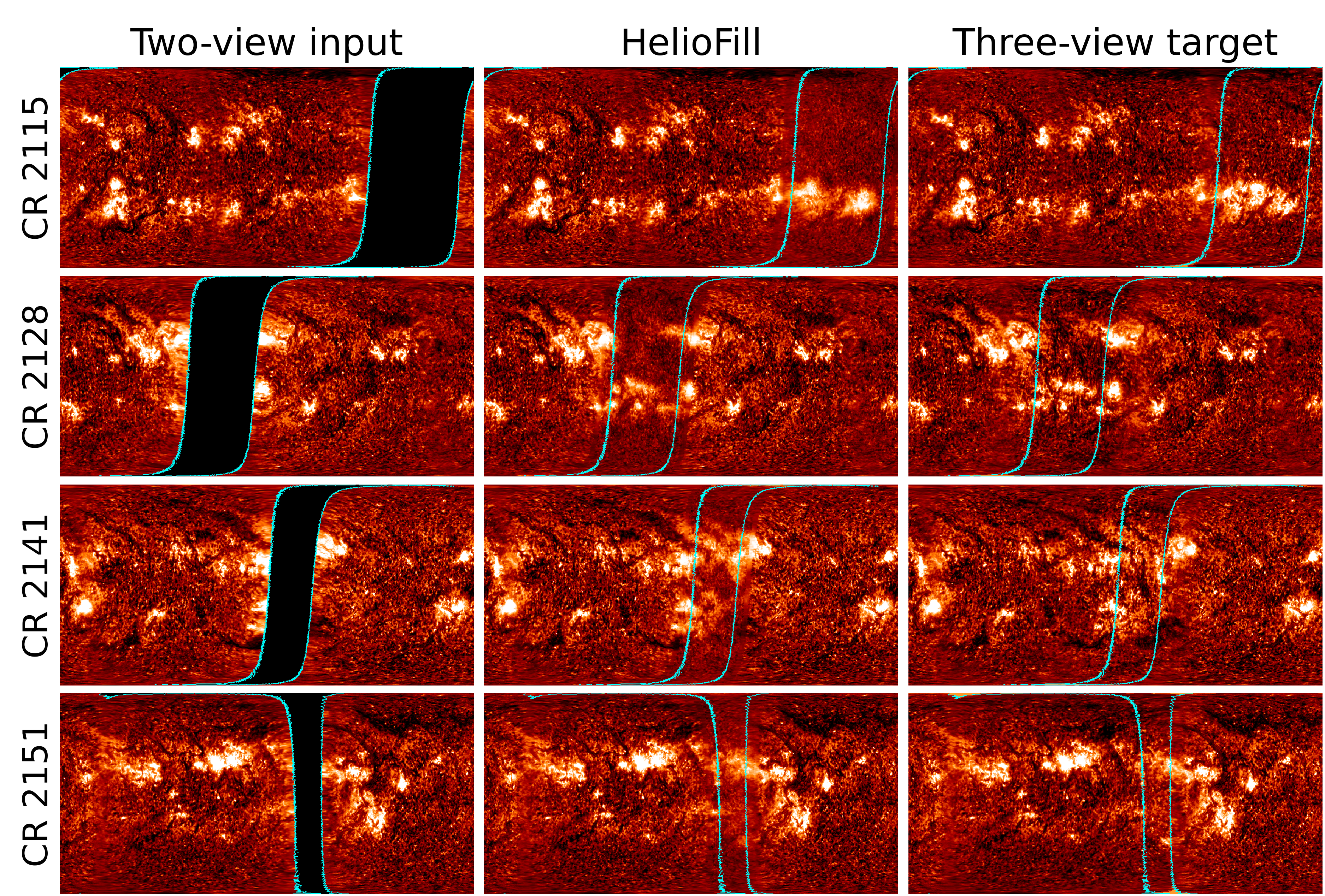}
\caption{HelioFill on one test map per held-out rotation (top to bottom: CR~2115, 2128, 2141, and 2151; 2011 October 14 04:00, 2012 September 18 23:00, 2013 September 04 18:00, and 2014 June 04 03:00\,UT), each the median of its rotation in HelioFill's correlation over the unobserved sector ($\rho_S=0.569$, $0.510$, $0.544$, and $0.604$; PSNR$_S$ $14.81$, $14.71$, $14.53$, and $16.50$\,dB). Left: two-view input (black: unobserved sector); middle: HelioFill; right: three-view target. Cyan: edge of $S$. All panels share one color scale. Figures~\ref{fig:baselines} and \ref{fig:stages} zoom in on the sector of these maps. The small-scale texture inside the estimated sector is generated by the refinement stage, which is not observed (Appendix~\ref{app:bands}).}
\label{fig:qual}
\end{figure}

\begin{figure}[!t]
\centering
\includegraphics[width=\linewidth]{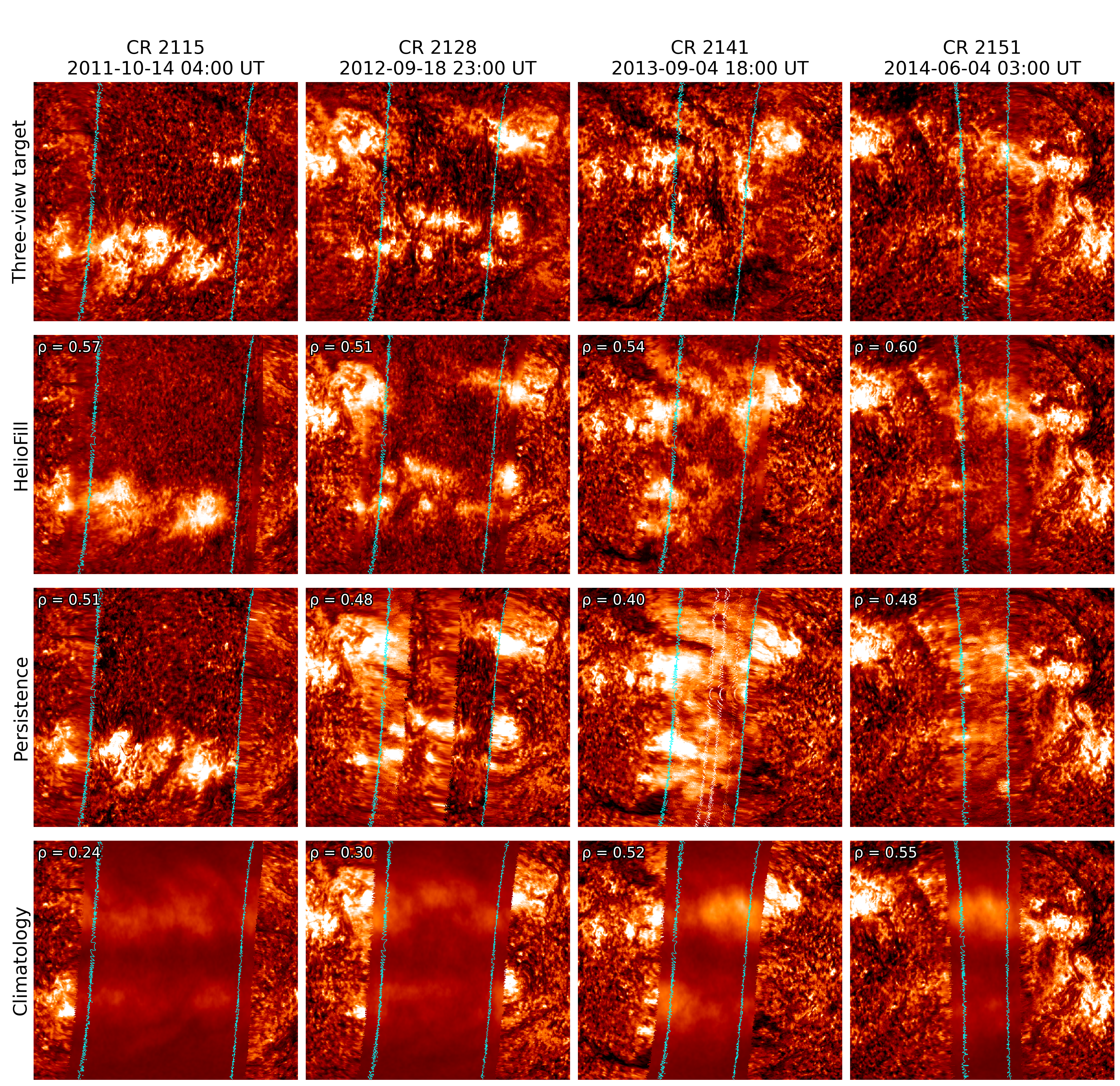}
\caption{The unobserved sector of the four maps of Figure~\ref{fig:qual} (columns), within $\pm45^{\circ}$ latitude, for the three-view target and the estimates of HelioFill, persistence, and climatology (rows), all on one color scale; cyan: edge of $S$; $\rho$: correlation with the target over $S$ (HelioFill $0.57$, $0.51$, $0.54$, and $0.60$; persistence $0.51$, $0.48$, $0.40$, and $0.48$; climatology $0.24$, $0.30$, $0.52$, and $0.55$). HelioFill places the active regions that persistence carries from the sector's last observation, without the seams and streaks of the persistence composite, and adds texture of realistic amplitude; climatology is a smooth latitudinal profile without active regions. The smoother strip 20--32\,px outside each edge of $S$ in HelioFill is the unrefined 12\,px band of Section~\ref{sec:pmrf}.}
\label{fig:baselines}
\end{figure}

\begin{figure}[!t]
\centering
\includegraphics[width=\linewidth]{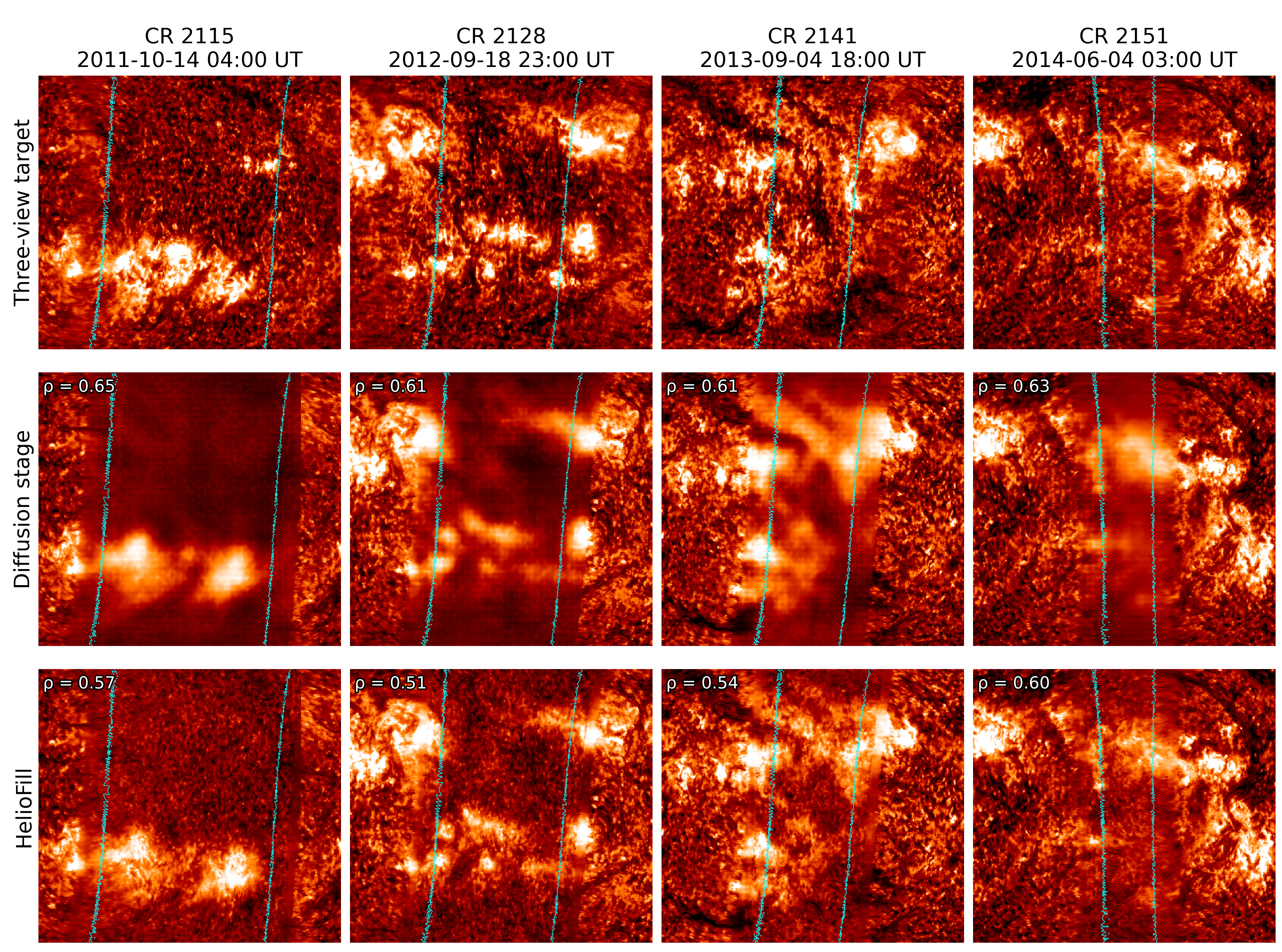}
\caption{The same sectors as Figure~\ref{fig:baselines} for the diffusion stage alone (harmonized sampling; $\rho=0.65$, $0.61$, $0.61$, and $0.63$) and HelioFill ($0.57$, $0.51$, $0.54$, and $0.60$), with the three-view target. The diffusion stage, a squared-error estimate, renders enclosed active regions as smooth patches at their approximate positions; the refinement stage adds small-scale texture of realistic amplitude that carries no positional information below $\approx$$5\fdg6$ (Appendix~\ref{app:bands}) and lowers the correlation.}
\label{fig:stages}
\end{figure}

We evaluate HelioFill against the retrained PixelHacker, nonlearning baselines, a prior-only control, and systematic ablations on full-Sun 304\,\AA\ CEA maps under the blocked protocol of Section~\ref{sec:train}. Table~\ref{tab:main} compares HelioFill with the baselines, the control, and its ablations on the 1330 test maps; Figures~\ref{fig:qual}--\ref{fig:stages} show one map per held-out rotation, each the median of its rotation in HelioFill's correlation over the unobserved sector. Over the sector, HelioFill's estimate correlates with the truth at $\rho=0.520$, above persistence ($+0.103\pm0.003$, higher on 97.9\% of maps), climatology ($+0.128\pm0.005$, 96.8\%), retrieval ($+0.295\pm0.008$, 99.8\%), and the prior-only control ($+0.340\pm0.008$, 99.7\%). In PSNR it exceeds persistence by $2.27\pm0.04$\,dB (99.8\%), retrieval by $3.72\pm0.08$\,dB (99.8\%), and climatology by $0.66\pm0.05$\,dB (83.1\%). Climatology, whose contrast over $S$ is $0.47$ of the target's ($0.79$ for HelioFill, $0.95$ for persistence), ranks above persistence in PSNR and below it in correlation: its structureless fill is the second-best of HelioFill and the baselines in PSNR while localizing almost no active regions (Figure~\ref{fig:baselines}; Section~\ref{sec:structure}). Persistence, a real image whose content is stale, has the opposite profile, texture of the right amplitude with seams and streaks in the wrong places.

The components of HelioFill act in different places. The history channels carry the agreement inside the sector: they raise the diffusion stage from $\rho=0.353$ to $0.546$ ($+0.192\pm0.005$, higher on 99.8\% of maps) and from $14.10$ to $15.70$\,dB ($+1.61\pm0.04$\,dB, 99.5\%), turning a model that trails climatology in both metrics ($-0.038\pm0.006$ in $\rho$, $-0.75\pm0.05$\,dB) into one that exceeds it. Harmonized sampling adds $0.029\pm0.002$ in $\rho$ (99.3\%) and $0.31\pm0.01$\,dB (91.9\%). The solar-specific modifications to PixelHacker act at the boundary: they improve PSNR by $1.77$\,dB in the part of $M$ outside $S$, but inside $S$ by only $0.05\pm0.03$\,dB and $0.031\pm0.004$ in $\rho$. The refinement stage then trades $0.055\pm0.001$ in $\rho$ (higher on 0.7\% of maps) and $0.50\pm0.02$\,dB for realistic texture, lowering LPIPS from $0.205$ to $0.182$ on 99.8\% of maps, the lowest value in the table (Figure~\ref{fig:stages}). The prior-only control shows that the agreement comes from the observations rather than the learned prior: blanking the conditioning of the model without history costs $0.174\pm0.008$ in $\rho$ (89.6\%) and $1.95\pm0.05$\,dB (99.4\%) over the sector while LPIPS even improves slightly (from $0.204$ to $0.194$), so the prior alone yields realistic-looking but misplaced texture.

\begin{table}[!t]
  \centering
  \caption{Agreement with the Three-view Target on the Blocked Test Set (304\,\AA, $n=1330$)}
  \label{tab:main}
  \small
  \setlength{\tabcolsep}{4pt}
  \begin{adjustbox}{width=\linewidth,center}
  \begin{tabular}{lcccccc}
  \toprule
  \textbf{Method} & $\boldsymbol{\rho}_S$ $\uparrow$ & \textbf{PSNR}$_S$ [dB] $\uparrow$ & \textbf{PSNR}$_M$ [dB] $\uparrow$ & \textbf{SSIM}$_S$ $\uparrow$ & \textbf{LPIPS} $\downarrow$ & \textbf{Seam} [$10^{-3}$] $\downarrow$ \\
  \midrule
  \multicolumn{7}{l}{\textit{Baselines and control}} \\
  Climatology & $0.392 \pm 0.159$ & $14.84 \pm 0.91$ & 14.97 & 0.085 & 0.223 & 4.30 \\
  Persistence & $0.416 \pm 0.156$ & $13.23 \pm 1.29$ & 13.50 & 0.048 & 0.198 & 3.91 \\
  Nearest-neighbor retrieval & $0.225 \pm 0.123$ & $11.78 \pm 1.55$ & 11.87 & 0.042 & 0.193 & 8.74 \\
  Prior-only control & $0.179 \pm 0.125$ & $12.15 \pm 0.74$ & 12.23 & 0.044 & 0.194 & 34.83 \\
  \addlinespace[0.25em]
  \multicolumn{7}{l}{\textit{Ablations}} \\
  Retrained PixelHacker & $0.322 \pm 0.166$ & $14.04 \pm 1.22$ & 13.94 & 0.075 & 0.205 & 4.00 \\
  Without history & $0.353 \pm 0.169$ & $14.10 \pm 1.08$ & 14.48 & 0.077 & 0.204 & 7.88 \\
  Diffusion stage, standard sampling & $0.546 \pm 0.171$ & $15.70 \pm 1.03$ & 15.88 & 0.101 & 0.201 & 5.42 \\
  Diffusion stage, harmonized sampling & $0.575 \pm 0.151$ & $16.01 \pm 0.90$ & 16.29 & 0.105 & 0.205 & 3.92 \\
  \addlinespace[0.25em]
  HelioFill & $0.520 \pm 0.159$ & $15.50 \pm 0.93$ & 15.74 & 0.082 & 0.182 & 4.51 \\
  \bottomrule
  \end{tabular}
  \end{adjustbox}

  \vspace{0.4em}
  \begin{minipage}{\linewidth}
  \footnotesize
  \textbf{Note.} Mean over test maps ($\pm$s.d.\ for $\rho_S$ and PSNR$_S$); seam is the median. $\rho_S$: Pearson correlation with the target over $S$. $S$: unobserved sector; $M$: masked region ($S$ widened by $7^{\circ}$ in longitude on each side). Each ablation disables one further component of HelioFill (Section~\ref{sec:exp-design}).
  \end{minipage}
\end{table}

\subsection{Equivalent Observation Age}\label{sec:freshness}

\begin{figure}[!t]
\centering
\includegraphics[width=\linewidth]{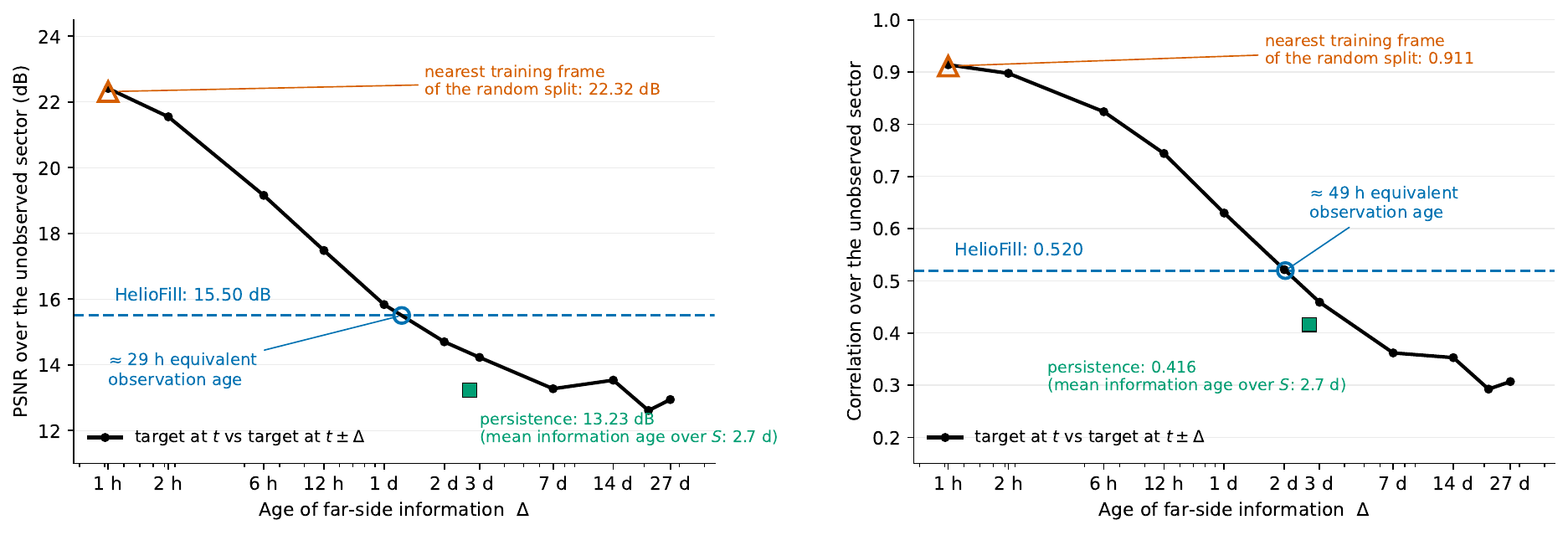}
\caption{Equivalent observation age. Black: PSNR (left) and correlation (right) over the unobserved sector of the target at $t$ scored against the target at $t\pm\Delta$. Dashed: HelioFill ($15.50$\,dB, $\rho=0.520$), which crosses the curves at $\approx$29\,hr and $\approx$49\,hr. Square: persistence at its mean information age. Triangle: the nearest training frame under a random split (median gap 1\,hr), the information that protocol exposed (Appendix~\ref{app:leakage}).}
\label{fig:freshness}
\end{figure}

Absolute scores are hard to interpret, so we calibrate them against the temporal self-consistency of the data: for each test map, the target over the unobserved sector at time $t$ is scored against the target at $t\pm\Delta$ (Figure~\ref{fig:freshness}; Table~\ref{tab:e9}). Self-consistency falls from $22.41$\,dB and $\rho=0.914$ at $\Delta=1$\,hr through $15.84$\,dB and $0.630$ at 1\,day to a recurrence plateau near 13\,dB and $0.3$ beyond a week. In PSNR over the sector, HelioFill matches a direct observation of the sector made about 29\,hr before or after the sample time; in correlation, one made about 49\,hr before or after. The diffusion stage alone, which scores higher on both, matches fresher observations, of about 22\,hr and 34\,hr. Its history input is of comparable age, yet copied directly, as persistence does, the same history reaches only $13.23$\,dB and $\rho=0.416$ at a mean information age over $S$ of 2.7\,days, against reference values of $14.36$\,dB and $0.477$ at that age; the SDO and STEREO-A views at time $t$ do not show the sector.

\subsection{Structure-aware Evaluation}\label{sec:structure}

Pixel-averaged scores can be dominated by quiet-Sun emission, so Table~\ref{tab:structure} evaluates active regions and coronal holes directly. HelioFill localizes ARs better than every baseline (Dice $0.449$, $+0.066\pm0.004$ over persistence, better on 85\% of maps) at 13\% of persistence's false-alarm rate. Separating true ARs by whether they touch the edge of $M$ isolates the underdetermined case. Without history, the diffusion stage detects boundary-crossing ARs at $0.244$ but enclosed ARs at only $0.073$, whereas persistence detects enclosed ARs at $0.636$: the best predictor of an enclosed structure is its last observation. History input removes the asymmetry: the diffusion stage detects enclosed ARs at $0.397$ and boundary-crossing ARs at $0.382$. Persistence, which copies structures verbatim, still detects more ARs in both groups and places boundary-crossing ones more precisely (median centroid errors $2.4$ against $3.9$\,px for HelioFill, and $2.8$ against $2.9$\,px for enclosed ones). The refinement stage raises enclosed-AR detection from $0.397$ to $0.480$ and CH Dice from $0.082$ to $0.110$, but lowers AR Dice by $0.014\pm0.002$ and more than doubles the AR false-alarm rate, so part of the structure it adds is misplaced. Coronal holes remain poorly reproduced by every method.

\begin{table}[!t]
  \centering
  \caption{Structure-aware Evaluation inside the Masked Region (Blocked Test Set, $n=1330$)}
  \label{tab:structure}
  \small
  \setlength{\tabcolsep}{4pt}
  \begin{adjustbox}{width=\linewidth,center}
  \begin{tabular}{lccccc}
  \toprule
   & \multicolumn{4}{c}{\textbf{Active Regions}} & \textbf{Coronal Holes} \\
  \cmidrule(lr){2-5}\cmidrule(lr){6-6}
  \textbf{Method} & Dice $\uparrow$ & False Alarm $\downarrow$ & Detection, Crossing $\uparrow$ & Detection, Enclosed $\uparrow$ & Dice $\uparrow$ \\
  \midrule
  Persistence & 0.384 & 0.116 & 0.450 & 0.645 & 0.080 \\
  Nearest-neighbor retrieval & 0.173 & 0.312 & 0.197 & 0.233 & 0.122 \\
  Climatology & 0.005 & 0.249 & 0.004 & 0.001 & 0.000 \\
  \addlinespace[0.25em]
  Without history & 0.197 & 0.030 & 0.243 & 0.073 & 0.065 \\
  Diffusion stage, harmonized sampling & 0.464 & 0.005 & 0.382 & 0.397 & 0.082 \\
  HelioFill & 0.449 & 0.015 & 0.384 & 0.480 & 0.110 \\
  \bottomrule
  \end{tabular}
  \end{adjustbox}

  \vspace{0.4em}
  \begin{minipage}{\linewidth}
  \footnotesize
  \textbf{Note.} Detection is split into the 5588 true AR components that touch the edge of $M$ or the map border and the 6317 that do not. False alarm is averaged over the maps with at least one predicted AR component. 
  \end{minipage}
\end{table}

\section{Far-side Validation Against Solar Orbiter}\label{sec:so}

All results above use 2011--2014 three-view mosaics. After 2014, the only direct full-disk EUV images of the unobserved sector come from the Extreme Ultraviolet Imager (EUI; \citealt{rochus2020eui}) aboard Solar Orbiter \citep{muller2020solo}. We reprojected 96 weekly EUI Full Sun Imager 304\,\AA\ images from 2022 March to 2025 September, taken while Solar Orbiter was $62^{\circ}$--$180^{\circ}$ in longitude from Earth, into the same Carrington frame and scored each estimate on the part of the unobserved sector that EUI sees away from its limb. Inputs are the current AIA and EUVI-A maps (no EUVI-A map was available for 16 of the 96 dates, whose input is AIA alone) and, as history, the preceding map archive, from AIA alone or, until 2023 March, AIA with EUVI-A (per-map mean information age on the scored region: median 6\,days, against 1.2\,days over $M$ on the test rotations); normalization uses observed pixels only. Both stages run exactly as on the test rotations: the mask handed to the model is the sector widened by 20\,px, observed pixels are copied back beyond 32\,px, and the refinement acts on the widened region (Appendix~\ref{app:pmrf} lists the values obtained with the unwidened mask). Because EUI and AIA differ in instrument and bandpass, the EUI data are calibrated affinely to AIA units on the day-side overlap, where the residual disagreement, scored with the same metrics but without the limb cut, is $16.38$\,dB (Table~\ref{tab:so}); PSNR against the range-compressed EUI maps therefore favors smooth estimates, and the Pearson correlation $\rho$ is the scale-free measure.

\begin{figure}[!t]
\centering
\includegraphics[width=\linewidth]{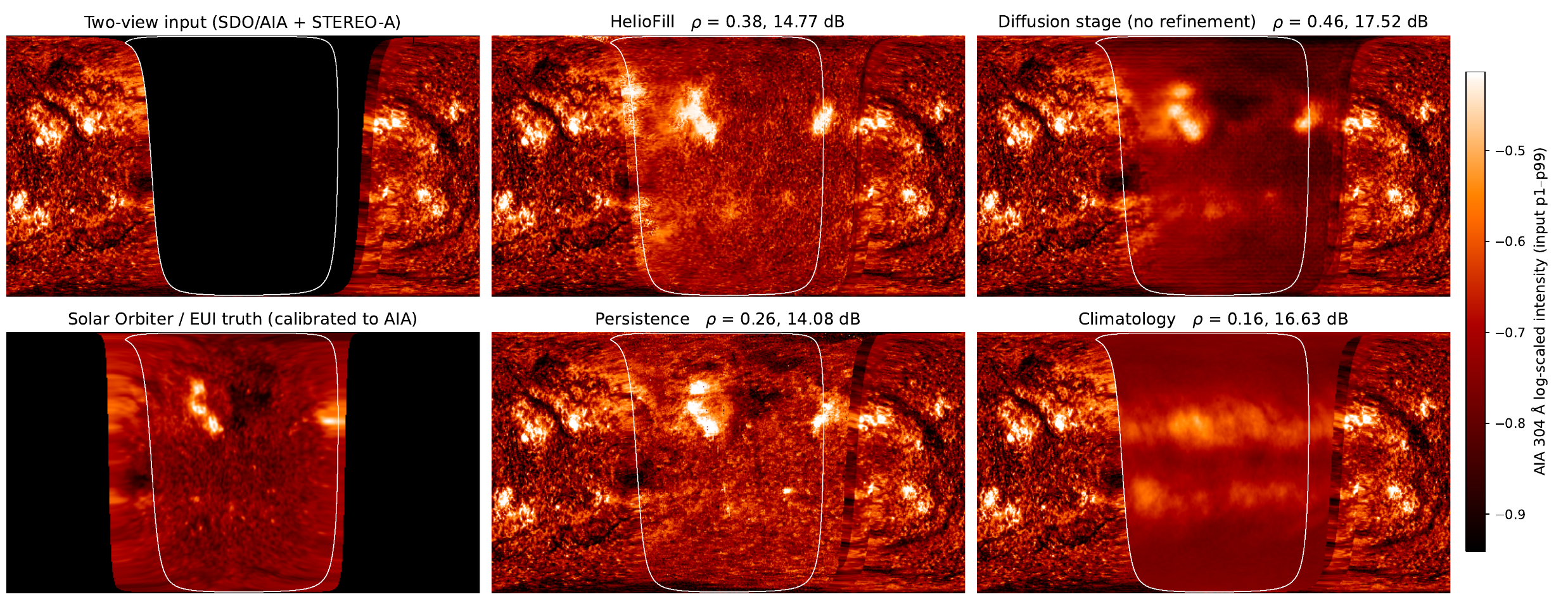}
\caption{Solar Orbiter validation for the lower of the two middle samples of HelioFill's correlation over the 96 (2022 August 12 12:00\,UT), rolled in longitude so that the scored region (white) is centered. Top: two-view input (black: observed by neither SDO nor STEREO-A), HelioFill ($\rho=0.38$), and the diffusion stage alone ($0.46$); bottom: EUI 304\,\AA\ truth calibrated to AIA units (black: not observed by EUI), persistence ($0.26$), and climatology ($0.16$), all on one color scale. In this sample, PSNR is $17.5$\,dB for the diffusion stage, $16.6$ for climatology, $14.8$ for HelioFill, and $14.1$ for persistence; Table~\ref{tab:so} gives the means over all 96 samples. Over all 96 samples, HelioFill correlates better than persistence on 61\% and than climatology on 81\%; the diffusion stage does so on 90\% and 93\%.}
\label{fig:so}
\end{figure}

HelioFill correlates with the far-side truth at $\rho=0.375$ (Table~\ref{tab:so}), above climatology ($+0.089\pm0.023$; higher on 81\% of samples) and level with persistence ($+0.010\pm0.013$; 61\%), which it exceeds in PSNR ($+0.43\pm0.23$\,dB) and SSIM ($+0.030\pm0.010$). History conditioning is what carries over to the post-2014 constellation: it raises the diffusion stage's correlation from $0.222$ to $0.460$ ($+0.238\pm0.027$; higher on 97\% of samples) and its PSNR by $2.24$\,dB (99\%); the sampler itself changes the far-side correlation by only $0.001$ ($0.4618$ with standard sampling, against $0.4604$). Climatology has the highest PSNR ($19.15$\,dB) because the calibrated EUI maps are range-compressed: its median contrast is $1.03$ of the truth's, against $1.68$ for the diffusion stage and $1.87$ for HelioFill. The refinement stage costs far-side correlation. The diffusion stage alone reaches $\rho=0.460$, above persistence on 90\% of samples ($+0.095\pm0.016$), and refinement lowers this by $0.085\pm0.007$, on 99\% of samples. It also brightens the scored region on every sample: the mean intensity rises by 13\% of the map's normalization range (the 1st--99th percentile range of the observed input) on average, against 1.3\% of the corresponding range over $S$ on the test rotations. Stochastic ($\eta=1$) standard sampling of the diffusion stage, an alternative route to texture, retains more far-side correlation ($\rho=0.391$; Appendix~\ref{app:pmrf}); the texture added by refinement is generated, not recovered (Appendix~\ref{app:bands}). Figure~\ref{fig:so} shows a median case.

\begin{deluxetable}{lccc}
\tablecaption{Far-side Validation Against Solar Orbiter/EUI at 304\,\AA\ ($N=96$, 2022--2025)\label{tab:so}}
\tablehead{\colhead{Method} & \colhead{$\rho$ $\uparrow$} & \colhead{PSNR [dB] $\uparrow$} & \colhead{SSIM $\uparrow$}}
\startdata
Climatology & $0.286 \pm 0.142$ & $19.15 \pm 2.01$ & $0.307 \pm 0.117$\\
Persistence & $0.365 \pm 0.127$ & $14.43 \pm 1.09$ & $0.287 \pm 0.104$\\
Without history & $0.222 \pm 0.133$ & $14.83 \pm 1.73$ & $0.252 \pm 0.114$\\
Diffusion stage, harmonized sampling & $0.460 \pm 0.159$ & $17.07 \pm 1.28$ & $0.402 \pm 0.129$\\
HelioFill & $0.375 \pm 0.139$ & $14.86 \pm 1.21$ & $0.318 \pm 0.113$\\
\hline
EUI versus AIA, day-side overlap & \nodata & $16.38 \pm 1.52$ & $0.351 \pm 0.188$\\
\enddata
\tablecomments{Mean $\pm$ s.d.\ over the 96 samples on the part of the unobserved sector that EUI sees away from its limb, in AIA log-intensity units. The \emph{Without history} row uses standard sampling and, like the two rows below it, the mask protocol of Section~\ref{sec:data}. The last row is the EUI--AIA disagreement where both observe, scored with the same metrics but without the limb cut; it is not an upper bound, because the calibrated EUI maps are range-compressed. On 13 of the 96 dates (2022 April--October), the AIA reference used for the calibration and for this row also contains EUVI-A data. Appendix~\ref{app:pmrf} records the earlier use of this set to compare refinement configurations. SSIM is computed as a single window over the scored region, unlike SSIM$_S$ in Table~\ref{tab:main}.}
\end{deluxetable}

\section{Discussion}\label{sec:discussion}

HelioFill demonstrates that domain-specific diffusion models can provide observationally plausible estimates of the solar sector left unobserved since the loss of STEREO-B. Its central design lesson is that the two available information sources are complementary and should be conditioned on jointly: the current two-view image constrains boundary-crossing structure, whereas the sector's rotation history constrains what the visible limb cannot. Physically, a structure that crosses the boundary of the sector is constrained by its visible context, whereas a structure fully enclosed in the sector appears in no current image, and its last observation, carried by the rotation history, is the best available information on it: without history, the diffusion stage detects enclosed active regions at only $0.073$, with history at $0.397$ (Section~\ref{sec:structure}). The same division carries over to the post-2014 constellation: on the far side observed by Solar Orbiter, the history input accounts for more than half of the diffusion stage's correlation with the truth (Section~\ref{sec:so}).

Supplying both, with the history censored during training, yields an estimator that exceeds every single-source baseline in correlation with the truth, in PSNR, and in active-region localization on the test rotations. On the far side observed by Solar Orbiter, the diffusion stage keeps a margin in correlation ($\rho=0.460$ against $0.365$ for persistence) and the refinement stage spends it: HelioFill's far-side correlation ($0.375$) is level with persistence, although it remains above persistence in PSNR and SSIM and above climatology in correlation. Because squared error favors estimates of reduced contrast, climatology ranks second among HelioFill and the baselines in PSNR on the test rotations although it localizes almost no active regions; the correlation over the sector, which is free of contrast, separates HelioFill from climatology and persistence on 97\%--98\% of the test maps and is the primary measure of agreement in this Letter. The components act in different places: the history channels carry the agreement inside the unobserved sector and harmonized sampling adds to it, the solar-specific training improves the rest of $M$, and the refinement stage adds realistic texture at a measured cost in correlation, on the test rotations and on the Solar Orbiter far side alike. 

Preserving spacecraft measurements with minimal perturbation is crucial for deployment. HelioFill retains the observed regions farther than 32\,px from the unobserved sector and confines generative estimation to the missing far-side sector and the 32\,px band around it, thereby extending coverage without overwriting measured data elsewhere. This design facilitates integration into existing processing pipelines while limiting changes to downstream products.

These capabilities result from design choices targeted specifically at scientific image inpainting. The latent-space formulation keeps inference compatible with operational cadence, and the solar-specific training improves agreement in the part of $M$ outside $S$ (Section~\ref{sec:fidelity}), and harmonized sampling lowers the seam (Table~\ref{tab:main}).

\textit{Limitations.} First, when an enclosed structure has emerged or changed since its last observation, the history input is stale by construction; the information-age map discloses this staleness but cannot remove it, and persistence still detects more individual active regions. Second, the morphology of enclosed structures is plausible rather than veridical: HelioFill's coherence with the truth is above chance at spatial scales down to $\approx$$5\fdg6$ and at chance below (Appendix~\ref{app:bands}), where its texture, realistic in amplitude, carries no positional information. The output must therefore not be read as a detector of individual hidden active regions or as a substitute for helioseismic far-side monitoring. Third, coronal holes are reproduced at low fidelity (CH Dice $0.11$). Fourth, the 12\,px band beyond the masked region is regenerated by the diffusion stage but not refined (Section~\ref{sec:pmrf}) and is visibly smoother than its surroundings.

Natural next steps include adding further coronal channels, such as 171, 193, and 211\,\AA, and extending the direct far-side validation against Solar Orbiter observations as they accumulate, and characterizing the consistency of successive estimates at operational cadence, as each new observation enters the history buffer. HelioFill therefore represents a concrete step toward AI-augmented continuous global solar monitoring and suggests broader applications of diffusion-based methods to observational gaps across heliophysics.

\begin{acknowledgments}
Firas B.A. acknowledges support from the KAUST Global Fellowship Program (Award No.\ RFS-KGFP2023-6044). R.D. acknowledges support from the FWO SB PhD fellowship (No.\ 1S66825N). For computational resources, this research used the Ibex cluster, managed by the KAUST Supercomputing Laboratory. This work was funded by KAUST via the student visiting program (grant No.\ BAS/1/1663-01-01).
S.P. is funded by the European Union (ERC-AdG agreement No.\ 101141362, Open SESAME). Views and opinions expressed are, however, those of the author(s) only and do not necessarily reflect those of the European Union or the European Research Council. Neither the European Union nor the granting authority can be held responsible for them. 
SDO/AIA data are courtesy of NASA/SDO and the AIA science team, and STEREO/SECCHI data of the SECCHI consortium led by the Naval Research Laboratory. Solar Orbiter is an ESA--NASA mission operated by ESA; the EUI instrument was built by CSL, IAS, MPS, MSSL/UCL, PMOD/WRC, ROB, and LCF/IO with funding from BELSPO/PRODEX, CNES, UKSA, BMWi through DLR, and SSO.
\end{acknowledgments}

\facilities{SDO (AIA), STEREO (SECCHI/EUVI), SolO (EUI)}

\software{aiapy \citep{Barnes2020}, SunPy \citep{sunpy_community2020}, SolarSoft \citep{SOLARSOFT}, PyTorch \citep{paszke2019pytorch}}

\appendix
\restartappendixnumbering

\section{Diffusion Stage: Training Settings}\label{app:model}

\textit{Optimization.} The GLA-augmented conditioned UNet is trained using noised latents generated by the DDPM forward process and optimized with the weighted $\epsilon$-prediction objective in Equation~\eqref{eq:ldiff}. The noise schedule is scaled-linear, with $\beta_t=1-\alpha_t$ rising from $8.5\times10^{-4}$ to $0.012$ over $T=1000$ steps. We use AdamW~\citep{loshchilov2019adamw} with learning rate $\eta=2\times10^{-5}$, $\beta_1=0.9$, and $\beta_2=0.999$, weight decay $0.01$, together with a 300-step linear warmup followed by cosine decay to 10\% of the peak over a 100-epoch schedule, with an effective batch of 64 maps (four A100 80\,GB GPUs, eight maps per GPU, two accumulation steps). A configured gradient-norm clip of 0.8 was applied after the optimizer step and therefore had no effect. EMA tracking with decay $0.9995$ is used to maintain a smoothed copy of the model parameters throughout training. 

\section{Temporal Leakage and Protocol Checks}\label{app:leakage}

\textit{Leakage under a random split.} An earlier random 70/15/15 split of this archive placed 99.1\% of test timestamps within 6\,hr of a training timestamp. For the submitted model, evaluated on that split ($17.33$\,dB over $M$ on its 1516 test maps, the value reported in the submitted version), PSNR over $M$ fell by $0.43$\,dB per doubling of the time gap to the nearest training sample, whereas PSNR over the observed region showed no trend. The model without history, retrained on the blocked split with the recipe of Appendix~\ref{app:model}, scores $14.48$\,dB. Interleaved training thus exposed roughly 3\,dB of temporal adjacency through the training set; the censored history channels recover $1.4$\,dB of it through the input.

\textit{Temporal self-consistency.} Table~\ref{tab:e9} lists the curves of Figure~\ref{fig:freshness}. Under the random split, the nearest training frame lay a median of 1.0\,hr from each blocked-test time (self-consistency $22.32$\,dB, $\rho=0.911$), and 69\% of blocked-test times were themselves in that training set. Self-consistency at 14\,days still exceeds the 27\,days value by $0.59$\,dB and by $0.046$ in correlation, so recurrence information survives half a rotation; the guard bands therefore span a full rotation.

\begin{deluxetable}{lccccc}
\tablecaption{Temporal Self-consistency Over the Unobserved Sector: PSNR and Correlation of the Target at $t$ Against the Target at $t\pm\Delta$ ($t$ in the Blocked Test Set, $t\pm\Delta$ the Nearest Map of the Archive within 3\,hr, Both Signs Pooled)\label{tab:e9}}
\tablehead{\colhead{$\Delta$} & \colhead{$n$ pairs} & \colhead{PSNR (dB)} & \colhead{Median} & \colhead{$\rho$} & \colhead{Median}}
\startdata
1\,hr & 2621 & $22.41\pm1.27$ & 22.55 & $0.914\pm0.060$ & 0.928 \\
2\,hr & 2623 & $21.55\pm1.11$ & 21.59 & $0.898\pm0.055$ & 0.912 \\
6\,hr & 2585 & $19.16\pm1.10$ & 19.15 & $0.824\pm0.077$ & 0.848 \\
12\,hr & 2514 & $17.48\pm0.95$ & 17.45 & $0.744\pm0.092$ & 0.776 \\
1\,day & 2399 & $15.84\pm0.83$ & 15.76 & $0.630\pm0.120$ & 0.672 \\
2\,days & 2205 & $14.70\pm0.68$ & 14.62 & $0.521\pm0.139$ & 0.564 \\
3\,days & 2195 & $14.23\pm0.56$ & 14.15 & $0.459\pm0.150$ & 0.489 \\
7\,days & 2057 & $13.27\pm0.71$ & 13.26 & $0.362\pm0.148$ & 0.369 \\
14\,days & 2037 & $13.53\pm0.91$ & 13.42 & $0.353\pm0.144$ & 0.375 \\
21\,days & 2012 & $12.61\pm0.84$ & 12.62 & $0.292\pm0.125$ & 0.296 \\
27\,days & 1927 & $12.94\pm1.03$ & 12.79 & $0.307\pm0.114$ & 0.302 \\
\hline
Random-split nearest training frame & 1318 & $22.32\pm1.37$ & 22.47 & $0.911\pm0.064$ & 0.927 \\
\enddata
\end{deluxetable}

\textit{Evaluation region.} The part of $M$ outside $S$ holds 28\% of $M$ on average (15\%--38\%) and is observed in the two-view input. It is not a copy task: the two-view input itself scores $14.40$\,dB there against the three-view target, below HelioFill's $16.41$\,dB. PSNR comparisons in Section~\ref{sec:fidelity} are therefore made over $S$, which no input spacecraft observes; structure metrics and LPIPS use $M$.

\textit{Normalization bounds.} The bounds of Equation~(\ref{eq:norm}) are percentiles of the full three-view target, which includes the unobserved sector. Rerunning both stages on the test maps with bounds computed from observed input pixels only changes HelioFill's PSNR over the unobserved sector by $-0.03\pm0.01$\,dB (over $M$, by $-0.03\pm0.01$\,dB), and LPIPS rises by less than $0.001$ ($0.182$ against $0.182$). 

\section{Band-resolved Amplitude and Coherence}\label{app:bands}
\setcounter{table}{0}

A single PSNR does not reveal at which spatial scales an estimate can be trusted. For every test map, we therefore take the largest rectangle inside the unobserved sector, window it, Fourier transform the estimate and the target, and group the Fourier plane into octave bands of spatial period. In each band, we report the amplitude ratio and the coherence (Table~\ref{tab:bands}),
\begin{equation}
T(k)=\sqrt{\frac{E_{\mathrm{est}}}{E_{\mathrm{gt}}}},
\qquad
\gamma(k)=\frac{\left|\left\langle \mathcal{F}_{\mathrm{est}}\,\overline{\mathcal{F}_{\mathrm{gt}}}\right\rangle\right|}
{\sqrt{E_{\mathrm{est}}E_{\mathrm{gt}}}} ,
\label{eq:bands}
\end{equation}
where $T$ measures whether structure of that size has the right \emph{amplitude} and $\gamma$ whether it is in the right \emph{place}; texture of the correct size in the wrong place has $T\approx1$ and $\gamma\approx0$. Both are ratios of sums over all 1330 maps, with 95\% intervals from 2000 bootstrap resamples over maps. The chance floor comes from phase-randomized surrogates of each target \citep{theiler1992testing}, which keep its amplitude spectrum, and the ceiling from a VAE round trip of the target, which bounds what the latent representation can carry; the target scored against itself returns $\gamma=1$ in every band.

\begin{deluxetable}{lccccccc}
\tablecaption{Band-resolved Amplitude $T$ and Coherence $\gamma$ in the Unobserved Sector (Blocked Test Set, $n=1330$)\label{tab:bands}}
\tablehead{\colhead{Band (px)} & \colhead{$T$ HelioFill} & \colhead{$\gamma$ HelioFill} & \colhead{$\gamma$ diffusion} & \colhead{$\gamma$ persist.} & \colhead{$\gamma$ clim.} & \colhead{Chance} & \colhead{Ceiling}}
\startdata
$>128$ & 0.908 & 0.892 & 0.899 & 0.801 & 0.419 & 0.484 & 1.000 \\
128--64 & 0.908 & 0.813 & 0.829 & 0.737 & 0.578 & 0.317 & 1.000 \\
64--32 & 0.812 & 0.563 & 0.604 & 0.421 & 0.231 & 0.180 & 0.999 \\
32--16 & 0.616 & 0.157 & 0.253 & 0.143 & 0.075 & 0.095 & 0.997 \\
\hline
16--8 & 0.614 & 0.048 & 0.045 & 0.059 & 0.043 & 0.053 & 0.985 \\
8--4 & 0.650 & 0.030 & 0.033 & 0.024 & 0.031 & 0.032 & 0.694 \\
4--2 & 0.658 & 0.016 & 0.012 & 0.011 & 0.017 & 0.017 & 0.032 \\
$<2$ & 0.680 & 0.020 & 0.017 & 0.017 & 0.020 & 0.023 & 0.018 \\
\enddata
\tablecomments{Bands are spatial periods on the $512\times1024$ grid ($16$\,px $\approx5\fdg6\approx70$\,Mm at the equator); the rule marks where positional information ends. ``Diffusion'' is the diffusion stage alone with harmonized sampling. Bootstrap 95\% intervals of $\gamma$: HelioFill $[0.150,0.167]$ at 32--16\,px and $[0.045,0.050]$ at 16--8\,px, diffusion stage $[0.242,0.263]$ and $[0.043,0.047]$, chance $[0.090,0.100]$ and $[0.050,0.056]$.}
\end{deluxetable}

Down to the 32--16\,px band, HelioFill is coherent with the target above the chance floor ($\gamma=0.157$ against $0.095$, nonoverlapping intervals), so structure larger than $\approx$$5\fdg6$ is localized rather than merely plausible; above 64\,px its coherence is within $0.02$ of the diffusion stage alone. From 16\,px down, coherence sits at the chance floor, although the refinement stage restores the amplitude of these scales to $T=0.61$--$0.68$, against $0.06$--$0.15$ for the diffusion stage alone: the added texture has realistic amplitude but no positional information, and it partly overwrites the positional information of the 32--16\,px band. The limit is not set by the latent representation, which transmits the target at $\gamma=0.985$ in the 16--8\,px band, nor removed by the history channels, whose small-scale content comes from an earlier rotation and lands at chance. Phase-randomizing each estimate instead of the target gives chance floors of $0.090$ and $0.045$ for HelioFill and $0.098$ and $0.049$ for the diffusion stage in the 32--16 and 16--8\,px bands, so neither conclusion depends on which field is randomized.

\section{Refinement Stage: Training, Selection, and Evaluation Record}\label{app:pmrf}

\textit{Training.} The velocity network is a convolutional UNet without attention (18.5\,M parameters, longitude-periodic padding) that receives $Z_u$, $\hat{y}$, $m$, $u$, and $\sigma_s$, and no timestamp, rotation number, or history input. It is trained for 27{,}000 steps, with $u$ drawn uniformly and $\sigma_s$ log-uniformly in $[0.03,0.35]$, on diffusion-stage estimates (harmonized sampling) and targets at the 2363 guard-band timestamps at least 7\,days from any validation or test timestamp. Because this separation is shorter than the full-rotation guard of the diffusion stage, we tested for an advantage from temporal proximity: on the 73 test maps nearest to these training timestamps (7.0--11.6\,days away, at the start or end of their rotation), the refinement changes PSNR over $M$ by $-0.436$\,dB, against $-0.550$\,dB on the other 1257 maps (7.6--22.4\,days away), a weak benefit ($0.11\pm0.06$\,dB) that is small compared with the decline itself.

\textit{Selection.} Because the flow loss is minimized by the conditional mean, selecting on it would favor the smoothest checkpoint; we instead select the checkpoint whose intensity dispersion inside $M$ best matches the target's on validation maps. On the validation rotations, 25 Euler steps with $\sigma_s=0.03$ gave the highest PSNR and AR Dice and the lowest AR false-alarm rate among $N_{\mathrm{E}}\in\{25,50,100\}$ and $\sigma_s\in\{0.03,0.12,0.25\}$ (and $\sigma_s=0.045$ at $N_{\mathrm{E}}=25$) with the first of the training configurations below; $\sigma_s=0.01$ and $0.02$, below the training range, gave higher PSNR and fewer AR false alarms but higher LPIPS and lower AR Dice. At this operating point, three training configurations were compared: the recipe above at 9000 steps, a network with 2.25 times the parameters at 9000 steps, and the adopted 27{,}000-step run, which is best on validation PSNR, LPIPS, and AR Dice. With harmonized sampling on the validation rotations, stochastic DDIM sampling of the diffusion stage ($\eta=1$), an alternative route to texture (Section~\ref{sec:so}), lowers LPIPS from $0.239$ to $0.212$, at a cost of $1.08$\,dB over $M$ and a rise in the AR false-alarm rate from $0.003$ to $0.019$.

\textit{Evaluation record.} Besides the configurations reported above, the following were scored on the test rotations: diffusion-stage checkpoints at epochs 15, 20, 25, 30, and 99, and a variant fine-tuned with lagged history; three further checkpoints each of the without-history and retrained-PixelHacker ablations (for the former, before its validation sweep); post hoc intensity-calibration and variance-matching operators; two nonlearned refinement controls; and the 9000-step refinement network. The Solar Orbiter set was first scored, for every learned configuration, with the unwidened sector mask handed to both stages (HelioFill $\rho=0.387$, diffusion stage $0.461$ with harmonized and $0.452$ with standard sampling, without history $0.144$, $\eta=1$ sampling $0.382$, epoch 99 $0.370$, lagged history $0.273$); the reported values rerun every configuration with the widened mask of Section~\ref{sec:data}, which changes HelioFill by $-0.012\pm0.006$ and the harmonized diffusion stage by less than $0.001$ (the without-history model by $+0.078\pm0.035$ and $\eta=1$ sampling by $+0.009\pm0.009$, to the $0.391$ of Section~\ref{sec:so}). The three refinement training configurations were also scored on this set, and their correlation there was a criterion in choosing HelioFill's configuration ($\rho=0.371$ for the 9000-step network, $0.283$ for the larger one, and $0.386$ for the adopted run, with standard sampling and the unwidened mask), so it is not an untouched test for the refinement stage.

\bibliography{HelioFill}{}
\bibliographystyle{aasjournalv7}

\end{document}